\documentclass[twocolumn]{aastex631}

\usepackage{xcolor}

\received{2025 March 20}
\revised{2025 May 28}
\accepted{2025 June 16}
\published{2025 July 22}

\submitjournal{ApJ}

\begin{document}

\title{Detecting Stellar Coronal Mass Ejections via Coronal Dimming in the Extreme Ultraviolet}

\correspondingauthor{James Paul Mason}
\email{james.mason@jhuapl.edu}

\author[0000-0002-3783-5509]{James Paul Mason}
\affiliation{Johns Hopkins University Applied Physics Laboratory, 11000 Johns Hopkins Rd, Laurel, MD 20723, USA}

\author[0000-0002-1176-3391]{Allison Youngblood}
\affiliation{Exoplanets and Stellar Astrophysics Laboratory, NASA Goddard Space Flight Center, Greenbelt, MD 20771, USA}

\author[0000-0002-1002-3674]{Kevin France}
\affiliation{Laboratory for Atmospheric and Space Physics, University of Colorado Boulder, Boulder, CO 80309}

\author[0000-0003-2073-002X]{Astrid M. Veronig}
\affiliation{University of Graz, Institute of Physics \& Kanzelh\"ohe Observators for Solar and Environmental Research, Universit\"atsplatz 5, 8010 Graz}

\author[0000-0002-9672-3873]{Meng Jin}
\affiliation{Lockheed Martin Solar and Astrophysics Lab (LMSAL), Palo Alto, CA 94304, USA}

\begin{abstract}

Stellar flares and coronal mass ejections (CMEs) can strip planetary atmospheres, reducing the potential habitability of terrestrial planets. While flares have been observed for decades, stellar CMEs remain elusive. Extreme ultraviolet (EUV) emissions are sensitive to both flares and CME-induced coronal dimming. We assess the detectability of stellar CME-induced EUV dimming events by adapting a known ``Sun-as-a-star'' dimming technique -- validated by the Solar Dynamics Observatory’s EUV Variability Experiment (EVE) -- to stellar conditions. We adapt the solar data to reflect a range of stellar intensities, accounting for intrinsic brightness, distance, and interstellar medium (ISM) attenuation. We generate synthetic light curves for two different missions: the legacy EUV Explorer (EUVE) and the proposed ESCAPE mission. Our results indicate that dimming detections are well within reach. EUVE's broadband imager was capable of detecting stellar CMEs -- albeit with limited spectral (temperature) resolution -- but that was not part of the observing plan. EUVE's spectroscopic survey lacked sufficient sensitivity for CME detections. Optimizing modern instrument design for this task would make the observation fully feasible. In this work, we present a tool to explore the stellar-CME detection parameter space. Our tool shows that an instrument with performance similar to ESCAPE, setting a 600-second integration period, and integrating the spectra into bands, any star with a X-ray flux $\geq 2.51 \times 10^{-12}$~erg~s$^{-1}$~cm$^{-2}$ should have a $\geq$3$\sigma$ detection even for a modest few-percent dimming profile, regardless of ISM attenuation. Such measurements would be crucial for understanding the space weather environments of exoplanet host stars and, ultimately, for evaluating planetary habitability.

\end{abstract}


\section{Introduction} \label{sec:intro}
The search for life beyond our solar system hinges on the habitability of exoplanets, which is largely shaped by the environment defined by their host stars. A star's radiative output not only determines whether a planet stays warm enough to support liquid water, but also whether its atmosphere can persist over long timescales. Extreme ultraviolet (EUV; $\sim$100-1210 \AA) radiation, in particular, drives atmospheric loss through a variety of thermal and non-thermal properties (e.g., \citealt{Lammer2003, Johnstone2015, Jakosky2018, Tu2015, Amerstorfer2017, Kulikov2006, Lichtenegger2016, Dong2017, Hazra2025}). Transient space weather phenomena, such as flares and coronal mass ejections (CMEs), can further accelerate atmospheric loss and both have strong signatures in the EUV. Flares are observed directly in EUV light, while CMEs can induce transient EUV dimming (e.g., \citealt{Reinard2008, Woods2012, Milligan2012, Mason2014, Mason2016, Dissauer2018, Dissauer2019}).

CME-induced coronal dimming appears in spectra as multiple emission lines around the ambient coronal temperature decreasing in intensity at the same time and persisting for several hours \citep{Mason2014}. \textit{The Sun exhibits} a sequence of iron emission lines in the range $\sim$170-300 \AA\ that cover the ambient coronal temperature range, spanning roughly 0.6-2.2 MK (\ion{Fe}{9} 171~\AA, \ion{Fe}{10} 177~\AA, \ion{Fe}{11} 180~\AA, \ion{Fe}{12} 195~\AA, \ion{Fe}{13} 202~\AA, \ion{Fe}{14} 211~\AA, \ion{Fe}{15} 284~\AA). The further away from the ambient coronal temperature the emission lines peak formation temperature is, the smaller the magnitude of the dimming (e.g., \citealt{Mason2014, Vanninathan2018}). Dimming depth scales with CME mass and the \textit{rate} of dimming (i.e., the slope of the light curve as the intensity decreases) scales with CME speed \cite[e.g.,][]{Mason2016, Dissauer2019, Jin2022}. However, the more energetic the CME, the more likely a flare is also present (\citealt{Schrijver2009} and references therein) which competes in the spectral light curve with the CME-induced dimming. Flares tend to rapidly (over seconds to minutes) heat plasma and then more gradually (over minutes to tens of minutes) cool; this manifests in EUV emission lines as a peak occurring early in hotter lines and later in cooler lines. As a result, these two competing effects can be distinguished with careful analysis if there aren't so many flares occurring that the dimming is completely obscured. These various time-dependent and temperature-dependent signatures also make clear that instrumentation designed to measure them should have good cadence and spectral coverage.

A significant challenge in observing stellar EUV is that the light is heavily attenuated by the interstellar medium (ISM). Despite this, flares are generally detectable due to their intensity, with typical EUV brightening factors ranging from 2x to 10x above baseline levels \citep{Tristan2023, Loyd2018, Loyd2018b, Audard2000}. In contrast, CME-induced coronal dimming \textit{lowers} an already faint signal, if only by a modest amount (typically 1-10\%{; \citealt{Mason2019} and references therein)\footnote{All observed \textit{solar} dimmings are in this modest range of magnitudes, but there is some evidence that other stars may produce ``extreme" dimmings on the order of 50\% \citep{Veronig2021}.}. This raises key questions: are CME-induced stellar dimmings detectable? How does detectability scale with star brightness and ISM attenuation? 

This paper addresses these questions by modeling the response of an observed solar EUV dimming event captured by the Solar Dynamics Observatory (SDO, \citealt{Pesnell2012}) EUV Variability Experiment (EVE; \citealt{Woods2012}) to mimic how it would appear if the event had occurred at a distant star. Section \ref{sec:data} provides more details on these Sun-as-a-star (spatially integrated) spectra. We simulate the observation of other stars by scaling the EVE solar measurements for distance, intrinsic brightness, and ISM attenuation (Section \ref{sec:adapting}). These adaptations enabled us to evaluate dimming in light curves from various emission lines, line combinations, and spectral bands. To model real-world instrument measurements, we then applied noise, adjusted spectral binning and time integration, and convolved with the effective area curves for two specific missions: the EUV Explorer (EUVE; \citealt{Bowyer1991}) and the recently proposed ESCAPE mission \citep{France2022}. Section \ref{sec:instrumentation} describes the instrumentation in more detail. We generated synthetic light curves for both EUVE and ESCAPE to examine the detectability of our representative dimming event. This approach allowed us to explore how each key parameter -- brightness, ISM attenuation, spectral binning, and time integration -- affected the detectability of the dimming event (Section \ref{sec:results}). 

\citet{Veronig2021} previously searched the EUVE archive for indications of coronal dimming events in the aftermath of stellar flares and found one on the young active star AB Dor, demonstrating the potential for detecting such events even with an instrument not optimized for this purpose. However, the event was much more extreme (much larger magnitude) than anything ever observed on the sun, likely due to an inherent selection effect resultant from the limited sensitivity of EUVE. In this work we study the size of the detectability space. Section \ref{sec:detection} describes our detection method and Section \ref{sec:results} then addresses the question, revealing that there is much promise for stellar dimming detection. In Section \ref{sec:discussion} we discuss the implications, limitations of the method, and alternative or complementary methods. Finally, Section \ref{sec:conclusion} reemphasizes the importance of these measurements in the search for life and outlines future prospects for building that capability. 

\section{Observational Data} \label{sec:data}
For decades, solar observations have been collected across multiple instruments, including EUV imagers, coronagraphs, and soft X-ray photometers. The resulting datasets provide a comprehensive view of eruptive events on the sun, our G2-type star. Recently, much effort has been devoted to understanding how CMEs might behave on other stars, primarily motivated by the burgeoning exoplanets field.

For example, stronger overlying coronal magnetic fields could inhibit or slow ejections \citep{Svestka1992, Drake2013, Drake2016, Alvarado-Gomez2018}, while stronger active region fields could lead to faster, more massive ejections (e.g., \citealt{Moschou2017}). To fully understand this plasma acceleration process, we must place solar behavior within the broader context of stars that exhibit different magnetic field strengths and topologies (e.g., \citealt{Strickert2024}).

EVE provides a valuable link between solar and stellar regimes. It observes the Sun but \textit{without} spatial resolution, collecting spectra analogous to stellar measurements. Thanks to its proximity, the Sun's EUV light is bright and unaffected by ISM attenuation. Additionally, EVE's absolute calibration \citep{Hock2010} gives us confidence in the reported irradiance (in W m$^{-2}$ nm$^{-1}$), which we can translate into photon counts over time at each specific EUV wavelength and then scale to any arbitrary distance. 

Table \ref{tab:eve_tech_specs} summarizes the relevant EVE instrument performance parameters. Since 2010, EVE has been continuously observing the Sun, capturing many coronal dimming events (e.g., \citealt{Woods2011, Mason2014, Mason2016, Mason2019, Veronig2021, Xu2024}). Many of these studies validate irradiance coronal dimming measurements against spatially-resolved observations and explore the relationship between dimming and CMEs. Figures \ref{fig:eve_light_curves} and \ref{fig:aia_dimming_by_feature} show an example of spectral and spatial coronal dimming for an event that occurred on 2011 August 4, studied by \citet{Mason2016}, with a corresponding 2080 km s$^{-1}$, $5.7 \times 10^{15}$ g CME. This event forms the basis for analysis in the present paper.

\begin{deluxetable}{c|cc|cc}
\tablecaption{Relevant SDO/EVE spectrograph channels (MEGS-A and MEGS-B) technical specifications; more detail can be found in \citet{Hock2010} 
\label{tab:eve_tech_specs}}
\tablehead{
\colhead{Parameter} & \colhead{MEGS-A} & \colhead{MEGS-B} \\
}
\startdata
    Years in operation & 2010-2014 & 2010-present \\
    Wavelength range & 60-370 \AA & 360-1060 \AA \\
    Spectral resolution & 1 \AA & 1 \AA \\
    Cadence & 10 s & 10 s \\
    Absolute Irradiance Accuracy & 20\% & 20\% \\
\enddata
\end{deluxetable}

\begin{figure*}
\plotone{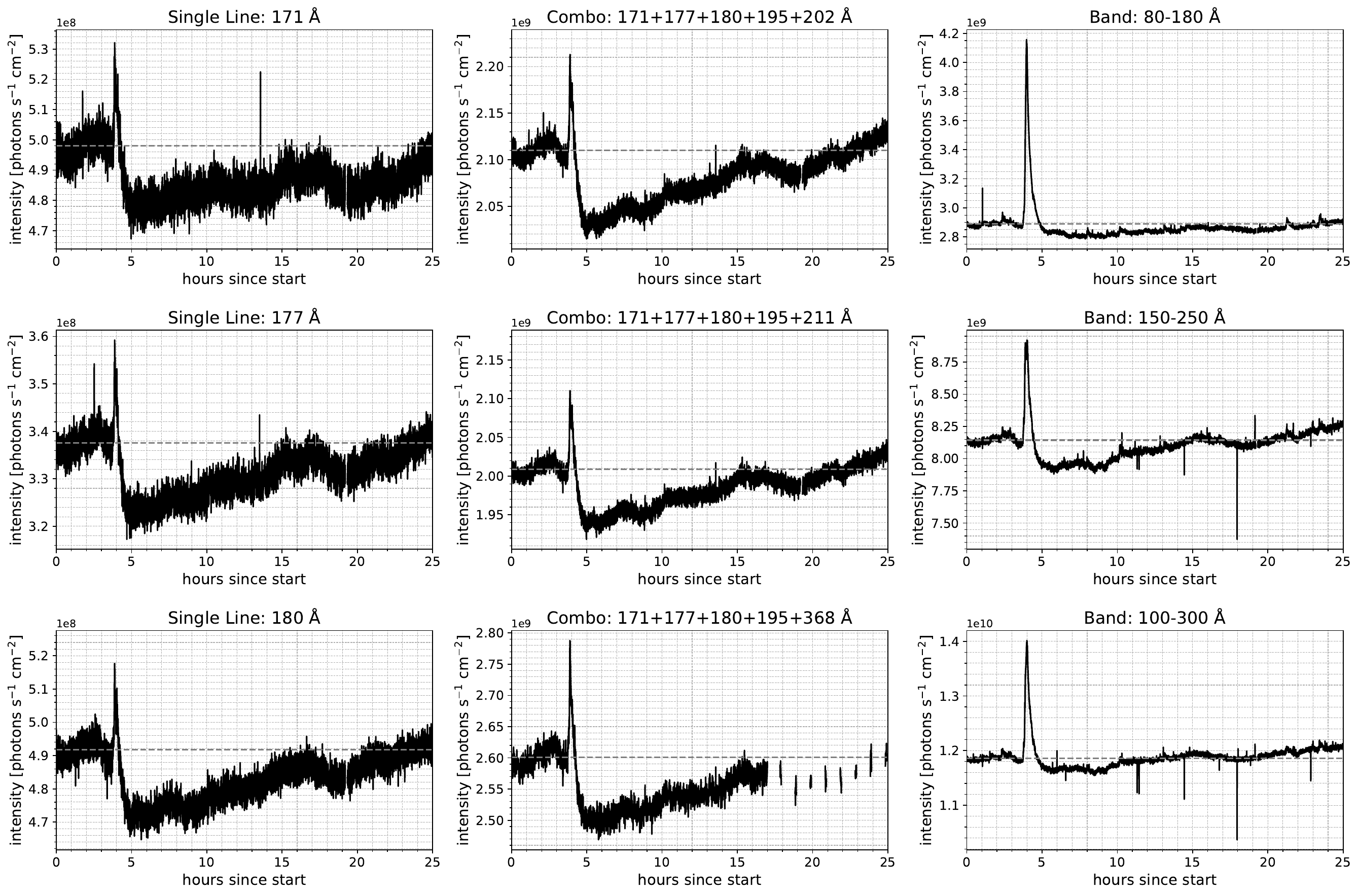}
\caption{Solar CME-induced coronal dimming event that occurred on 2011-08-04 as observed by EVE. The first column presents selected individual emission lines, the second column presents co-adds of 5 emission lines\footnote{The bottom middle plot has data gaps starting around 17 hours because the 350-1050 \AA\ channel went into a 5-minutes-per-hour observation mode to minimize long-term instrument degradation.}, and the third column presents spectrally integrated broad bands. Each row presents a different example of the column's category; many more curves were analyzed but this set is representative. Dashed horizontal lines indicate the pre-eruption baseline. The dimming shown here runs from hours $\sim$4-25 and is more readily observable in some wavelengths (those shown) than some others that are not as sensitive to dimming (not shown). 
\label{fig:eve_light_curves}}
\end{figure*}

\begin{figure*}
\plotone{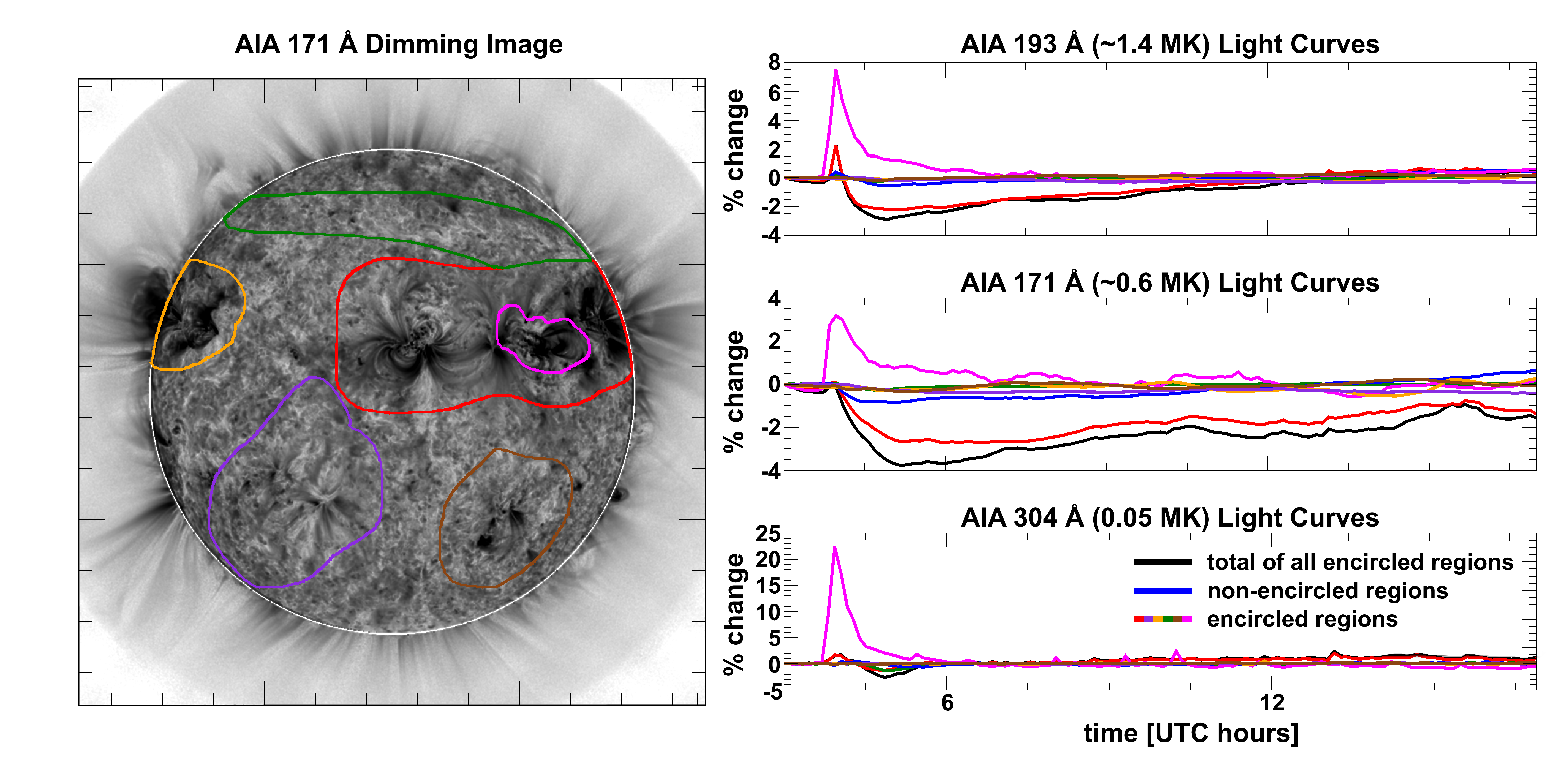}
\caption{The 2011-08-04 coronal dimming event as spatially resolved by SDO Atmospheric Imaging Assembly (AIA; \citealt{Lemen2012}). The colored contour overlays surround the dimming region of interest (red) while excluding the flaring region which is indicated in pink, and the other major features on the disk at the time that may have responded sympathetically to the eruptive event (filament in green, active regions in remaining colors). Light curves on the right are color coded correspondingly; each of the three vertically stacked plots corresponds to a different AIA bandpass. The light curve units have been converted to a percent change since the native units are uncalibrated DN and thus not directly comparable with the units shown in other figures. 
\label{fig:aia_dimming_by_feature}}
\end{figure*}

\section{Adapting Solar Observations to the Stellar Context} \label{sec:adapting}

To evaluate the detectability of CME-induced coronal dimming on other stars, we need to adapt solar EUV observations to the stellar context. Three parameters affect the apparent brightness: the intrinsic brightness of the star (i.e., luminosity), its distance, and the ISM column density of attenuating material along the sight-line\footnote{Atomic gases dominate attenuation within about 100 pc; in this local bubble, dust densities are low due to high radiation pressure \citep{Lehner2003}}. Luminosity and distance can be combined into a single value, flux, via the relationship $F = \frac{L}{4\pi d^2}$, where $F$ is the flux in units of erg~s$^{-1}$~cm$^{-2}$, $L$ is luminosity, and $d$ is distance. Values for the EUV luminosity or flux of stars other than the Sun are sparse due to the scarcity of direct observations: only EUVE has made such measurements and its limited effective area restricted the number of stellar luminosity measurements possible. However, extensive soft X-ray (SXR) flux measurements exist for both the Sun—collected over decades by instruments like GOES/XRS—and other stars observed with ROSAT \citep{Trumper1982}, Chandra \citep{Weisskopf2002}, and XMM-Newton \citep{derHerder2001}. Since SXR and EUV emissions are adjacent in the electromagnetic spectrum and are closely linked through coronal processes, SXR measurements serve as a reasonable proxy for EUV emissions \citep{Ribas2005, SanzForcada2011, Toriumi2022}.

\citet{Judge2003} determined that the Sun's average SXR luminosity over the solar cycle is approximately $10^{27.6}$erg~s$^{-1}$, corresponding to a flux of 1.415~erg~s$^{-1}$cm$^{-2}$ at a distance of 1 au. Measured X-ray fluxes of stars out to 50 pc range from approximately $7.4 \times 10^{-14}$ to $1.7 \times 10^{-10}$~erg~s$^{-1}$~cm$^{-2}$ \citep{Freund2022}. Because these fluxes span several orders of magnitude, we present them on a base-10 logarithmic scale (log$_{10}$\textit{F}) in subsequent figures. By taking the ratio of these measured solar and stellar flux values, we can scale solar observations to other stars, effectively accounting for both distance and intrinsic brightness.

We estimate the interstellar opacity along the line of sight to any given target by modeling the gas with a single neutral hydrogen column density (N(\ion{H}{1})), and assuming an ionization fraction of helium (0.6) and neutral hydrogen to helium ratios (0.08) from the observed local ISM values from \citet{Dupuis1995}.  N(HI) values are taken from the hundreds of interstellar sightlines with direct measurements from Hubble Space Telescope Ly$\alpha$ observations or extrapolated projections from three-dimensional maps (right ascension, declination, and distance; see \citet{Youngblood2025}).  The atomic opacity is converted into an attenuation curve for a given sightline that is applied to calculate the transmission of the stellar flux.

Where Figure \ref{fig:eve_light_curves} showed the solar intensity, Figure \ref{fig:ism_light_curves} is formatted the same way but with stellar intensities which are much lower, as expected. The dimming signal remains visible with reduced magnitude as the ISM attenuation increases. Whether or not this signal is detectable depends on the performance of the instrument measuring it, which is the subject of the next section. 

\begin{figure*}
\plotone{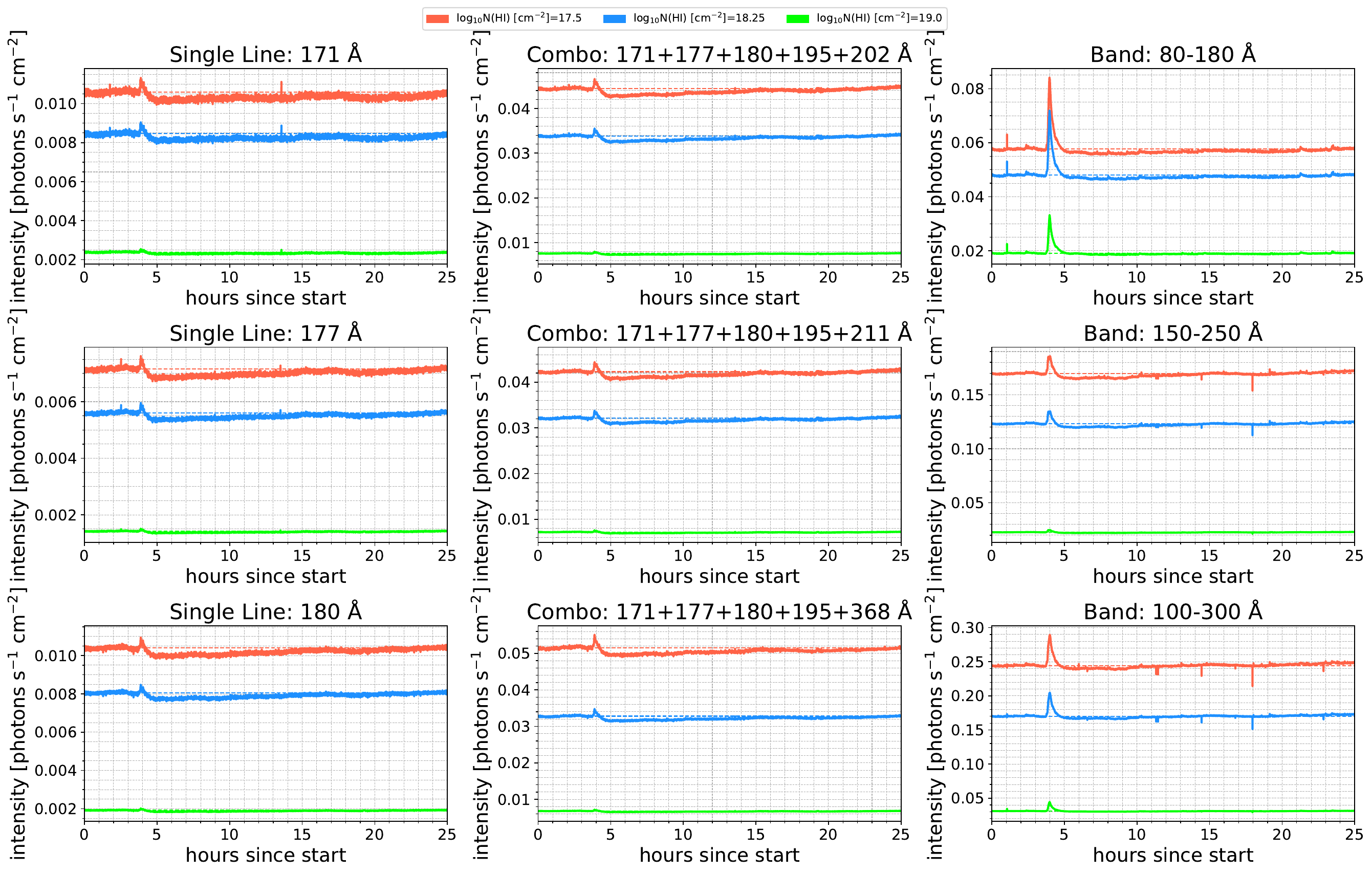}
\caption{Same format as Figure \ref{fig:eve_light_curves} but for a star with a log$_{10}$\textit{F} of -10.5 (i.e, $\sim$10$^{10}\times$ less flux than the solar reference of log$_{10}\textit{F} \approx$0.15). Each colored line represents a particular column density of \ion{H}{1} contributing to ISM attenuation. For context, the nearest stars have \ion{H}{1} absorbing columns of 10$^{17.6}$ cm$^{-2}$ and integrated \ion{H}{1} column densities rarely exceed 10$^{18.4}$ cm$^{-2}$ within 30 pc of the Sun \citep{Wood2005}. Instrument performance is not included here, but is included in the figures referenced in Section \ref{sec:instrumentation} and onward.
\label{fig:ism_light_curves}}
\end{figure*}

\section{Instrumentation} \label{sec:instrumentation}

Table \ref{tab:euve_escape_tech_specs} summarizes the instrument technical specifications for EUVE and ESCAPE, Figure \ref{fig:effective_area} shows their effective area curves, and Figures \ref{fig:euve_light_curves}-\ref{fig:escape_light_curves} show the corresponding application to the Figure \ref{fig:ism_light_curves} light curves. EUVE was a NASA Medium-Class Explorer (MIDEX) operated from 1992 to 2001, conducting both an all-sky survey and longer-duration stare campaigns~\citep{Malina1994}. The all-sky survey was conducted over a 6-month period with scanning telescopes in 4 photometric broad passbands, each day collecting new circular slices of the sky as the spacecraft rotated 3 times per orbit; as a result the observation time per target was quite short and we don't include it in the subsequent analysis presented here. The longer-duration stare campaigns were conducted with the Deep Survey / Spectrometer (DS/S), which itself was composed of an imager segment (effective area published in \citealt{Malina1994, Sirk1997}) and 3 separate spectrometer channels (effective areas described in \citealt{Hettrick1985}). EUVE's targets included white dwarfs, O- and B-type stars, and main-sequence stars with distances ranging from $\sim$1 to 150 pc.

\begin{deluxetable*}{ccccc} 
\tablecaption{Relevant EUVE and ESCAPE technical specifications; more detail can be found in \citet{Welsh1990, Malina1994, Drake1995} and \citet{France2022} \label{tab:euve_escape_tech_specs}}
\tablehead{
\colhead{Parameter} & 
\colhead{EUVE all-sky} &
\colhead{EUVE DS/S} & 
\colhead{EUVE DS/S} & 
\colhead{ESCAPE} \\
\colhead{} & 
\colhead{scanning telescopes} & 
\colhead{imager} & 
\colhead{spectrometers (3)} & 
\colhead{} \\
\colhead{} &
\colhead{(Scanners A/B, C)} & 
\colhead{} & 
\colhead{(SW, MW, LW)\tablenotemark{a}} & 
\colhead{} \\
}
\startdata
    Years in operation & 1992-2001 & 1992-2001 & 1992-2001 & future (proposed) \\
    Wavelength range [\AA] & 58~--~742 & 67~--~364 & 70~--~760 & 80~--~1650 \\
    Spectral resolution [\AA] & 4 photometric bands & 3 photometric bands & 0.5~--~2.0 & 0.6~--~1.0 \\
    Peak effective area [cm$^{2}$] & 12 & 28 & 0.5~--~2.0 & 100 \\
    Figure \ref{fig:effective_area} plot & N/A & blue & red & black
\enddata
\tablenotetext{a}{SW, MW, LW stand for short, medium, and long wavelength, respectively. These are also sometimes referred to in the literature as Channel A, B, and C \citep{Hettrick1985}. The numbers presented here and plotted in Figure \ref{fig:effective_area} are the merge/sum of all three channels.}
\end{deluxetable*}

ESCAPE was recently proposed as a NASA Small Explorer (SMEX) that proceeded into a competitive Phase A study~\citep{France2022}. Unlike EUVE, ESCAPE is a purely spectroscopic survey instrument, planning to conduct both broad and focused surveys to characterize a wide range of targets and also capture temporal dynamics. Both surveys will be conducted with the same instrument -- the only difference being the duration spent observing each target. The ESCAPE mission is focused on supporting the search for habitable worlds and as such the targets are mainly Sun-like stars (F-, G-, K- type) and M dwarfs with distances ranging from $\sim$1 to 50 pc. The main advancement in EUV astrophysics instrumentation that ESCAPE makes is in sensitivity (Figure \ref{fig:effective_area}); the effective area is 1-2 orders of magnitude higher than EUVE's DS/S imager and 2x-5000x higher than EUVE's DS/S spectrometers. Effective area of at least $\sim$10 cm$^{2}$ is necessary for \textit{extreme} stellar dimming detectability and $>$~40cm$^{2}$ with spectral resolution $<$~2 ~\AA\ is required to detect solar-like dimming on nearby stars (this will be demonstrated in Section \ref{sec:results}). Section \ref{sec:detection} describes how we perform that detection.

\begin{figure}
\plotone{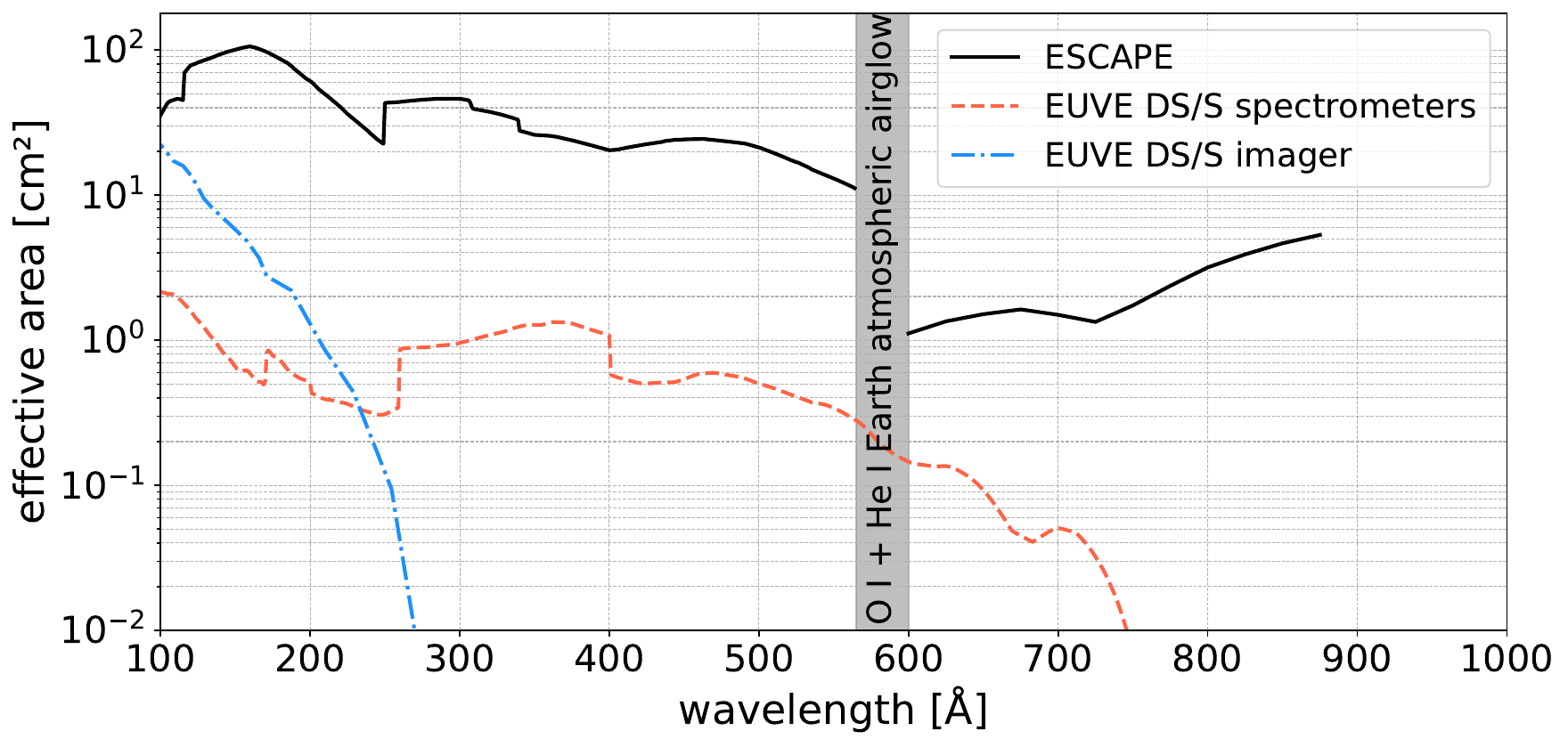}
\caption{Effective areas for the EUVE and ESCAPE missions. For this comparison we have summed all 3 EUVE wavelength channels (short, middle, long) since they were measured simultaneously and have some overlap. ESCAPE also has an FUV channel (1250-1800 \AA) that is not shown here. The Earth atmospheric airglow is a spectral region where there is significant EUV emission that interferes with measurements from any low Earth orbit platform.
\label{fig:effective_area}}
\end{figure}

\section{Dimming Detection} \label{sec:detection}

Figure \ref{fig:escape_single_light_curve} shows an example light curve with uncertainties and with colored highlights indicating a baseline level and the region of the curve used to calculate a dimming depth. The baseline level is the median intensity spanning the 4 hours prior to the event (highlighted in blue), which is starkly indicated by a flare then leading into the dimming. In this example, we spectrally integrated over the five emission lines indicated and then summed their intensities, accounting for the propagation of uncertainty through those manipulations.

\begin{figure*}
\plotone{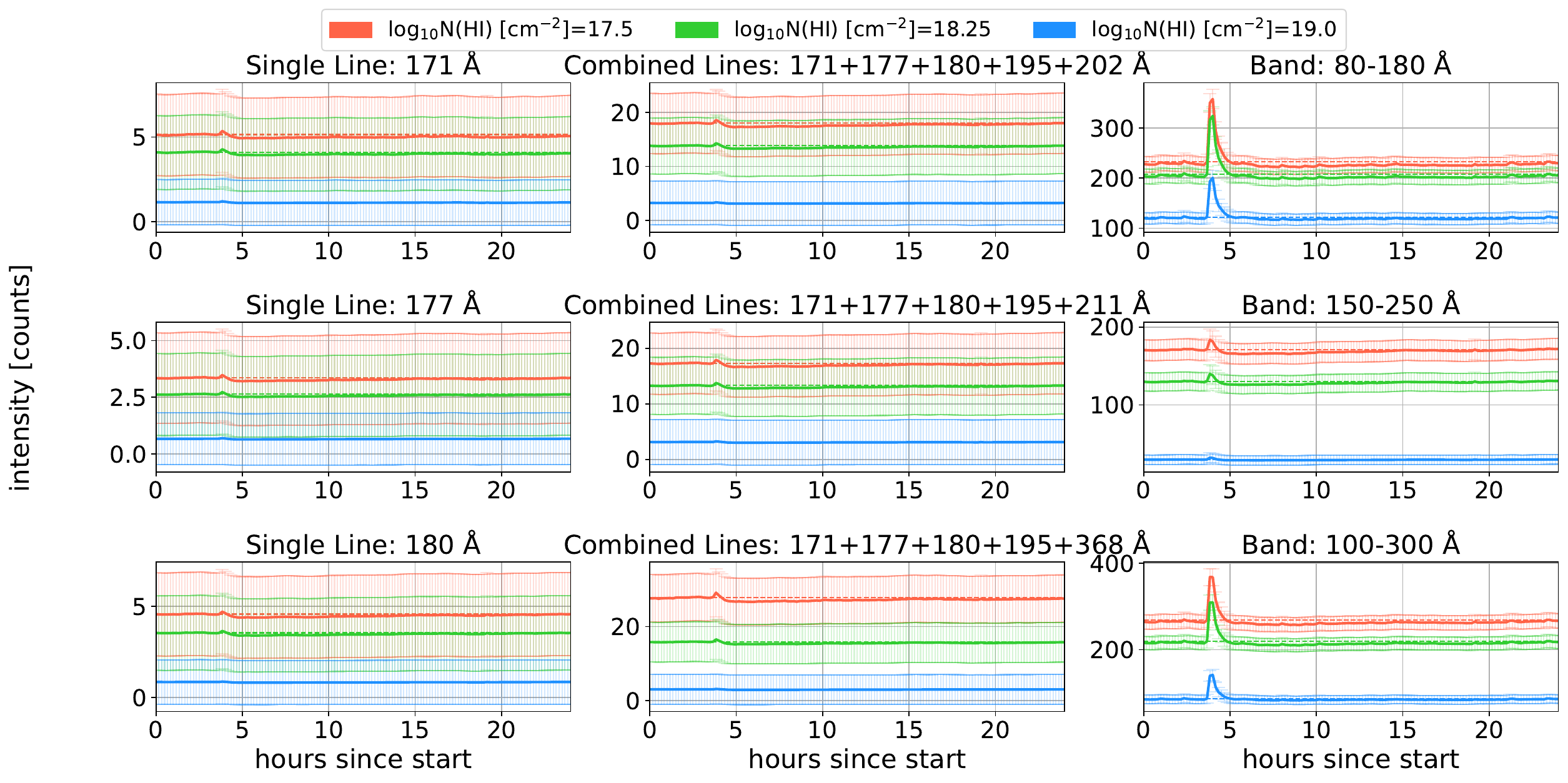}
\caption{Same format as Figure \ref{fig:ism_light_curves} but with the corresponding effective areas of EUVE applied (DS/S spectrometers effective area for the emission lines shown in the left two columns and DS/S imager effective area for the spectral bands in the right column), assuming a 600-second integration time. Instrument noise is also included and represented as error bars in the same color as the corresponding plot line; those errors are so large in the first two columns because the counts are so low.
\label{fig:euve_light_curves}}
\end{figure*}

\begin{figure*}
\plotone{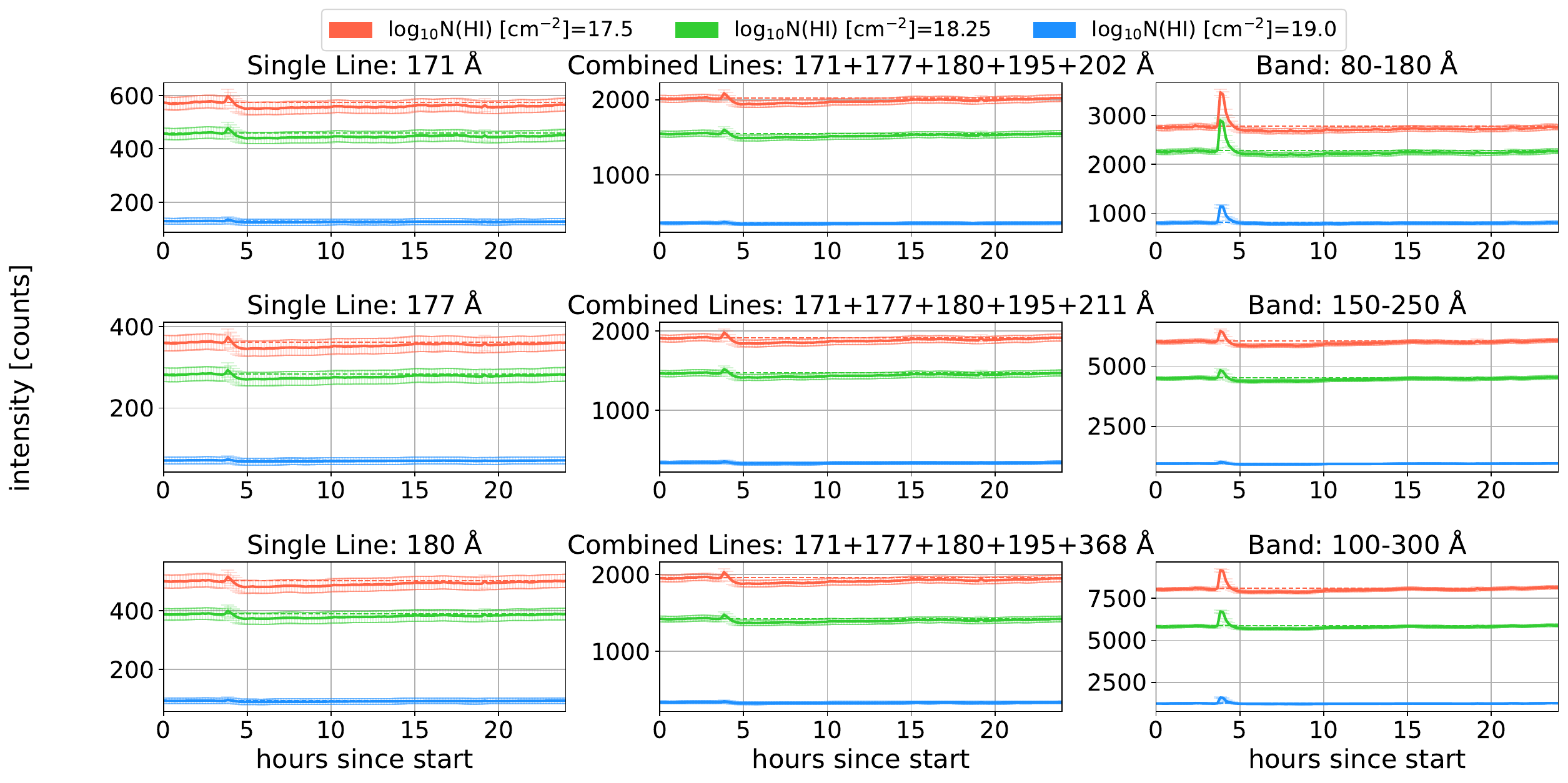}
\caption{Same format as Figure \ref{fig:euve_light_curves} but with the effective area and noise of ESCAPE applied. Note that noise is dramatically more constrained here due to the higher counts. See Figure \ref{fig:escape_single_light_curve} for an example zoomed in on a single light curve.
\label{fig:escape_light_curves}}
\end{figure*}

While the depth of the dimming could be calculated as simply the absolute minimum value within some window, that would have an uncertainty on the same scale as the individual points. Instead, we apply the following algorithm: 

\begin{enumerate}
    \item \textbf{Identify the absolute minimum}: find the absolute minimum in the window between the intensity dropping and returning to baseline level
    \item \textbf{Determine the halfway point}: calculate the intensity halfway between the baseline and the absolute minimum
    \item \textbf{Define the trough region}: select all intensities below this halfway point to constitute the trough (highlighted in red in Figure \ref{fig:escape_single_light_curve}, corresponding to a $\sim$9-hour period). 
\end{enumerate}

With the trough identified, we aim to calculate a singular dimming depth value. We can take advantage of the fact that we have many intensity points in the trough to reduce the uncertainty of the average value by taking the error-weighted mean intensity; in other words, we have alleviated the sensitivity to random variation. We calculate the weighted mean intensity, $\overline{I}$, and its associated error, $\sigma_{\overline{I}}$, as 

\begin{equation}
\overline{I}= \frac{\sum \frac{I}{\sigma^2}}{\sum \frac{1}{\sigma^2}}, \quad
\sigma_{\overline{I}} = \sqrt{\frac{1}{\sum \frac{1}{\sigma^2}}}.
\end{equation}

\noindent where \textit{I} is the observed intensity over time and $\sigma$ is the corresponding uncertainty (i.e., error bar of the observed intensity). These values are computed at two key intervals: the baseline and trough periods defined by the algorithm above and highlighted in Figure \ref{fig:escape_single_light_curve}. Finally, we can compute the depth itself, which is the change from the baseline intensity. We simultaneously convert to a percentage difference by dividing by the baseline intensity. 

\begin{equation}
\text{depth} [\%] = 100 \cdot \frac{\overline{I}_\mathrm{baseline} - \overline{I}_\mathrm{trough}}{\overline{I}_\mathrm{baseline}}
\end{equation}

We propagate the uncertainty through this calculation as well, resulting in 

\begin{equation}
\sigma_\mathrm{depth} [\%] = 100 \cdot \sqrt{\sigma_{\overline{I}_\mathrm{trough}}^2 \left(\frac{1}{\overline{I}_\mathrm{baseline}}\right)^2 + \sigma_{\overline{I}_\mathrm{baseline}}^2 \left(\frac{\overline{I}_\mathrm{trough}}{\overline{I}_\mathrm{baseline}^2}\right)^2}
\end{equation}

\noindent where $\sigma_{\overline{I}_\mathrm{baseline}}$ is the uncertainty of the weighted mean intensity for the baseline. This equation makes explicit the advantage of including multiple points in our depth calculation: the uncertainty of the depth is significantly reduced compared to calculating the depth between a single baseline and trough point, which is also a \textit{more accurate} assessment of the depth uncertainty. 

To determine whether a dimming event is detectable in any given light curve, we compare the magnitude of the depth (signal) with its uncertainty (noise). Simply dividing the two yields the signal to noise ratio, or what we will refer to as detection significance. For instance, a depth of 3\% with uncertainty of 1\% would be a 3$\sigma$ detection. Put another way, detection significance rises if either the underlying event is larger or if we can reduce the error bars on our measurement of it. Section \ref{sec:results} will show how the significance of detection scales with the key parameters of interest.

\begin{figure}
\plotone{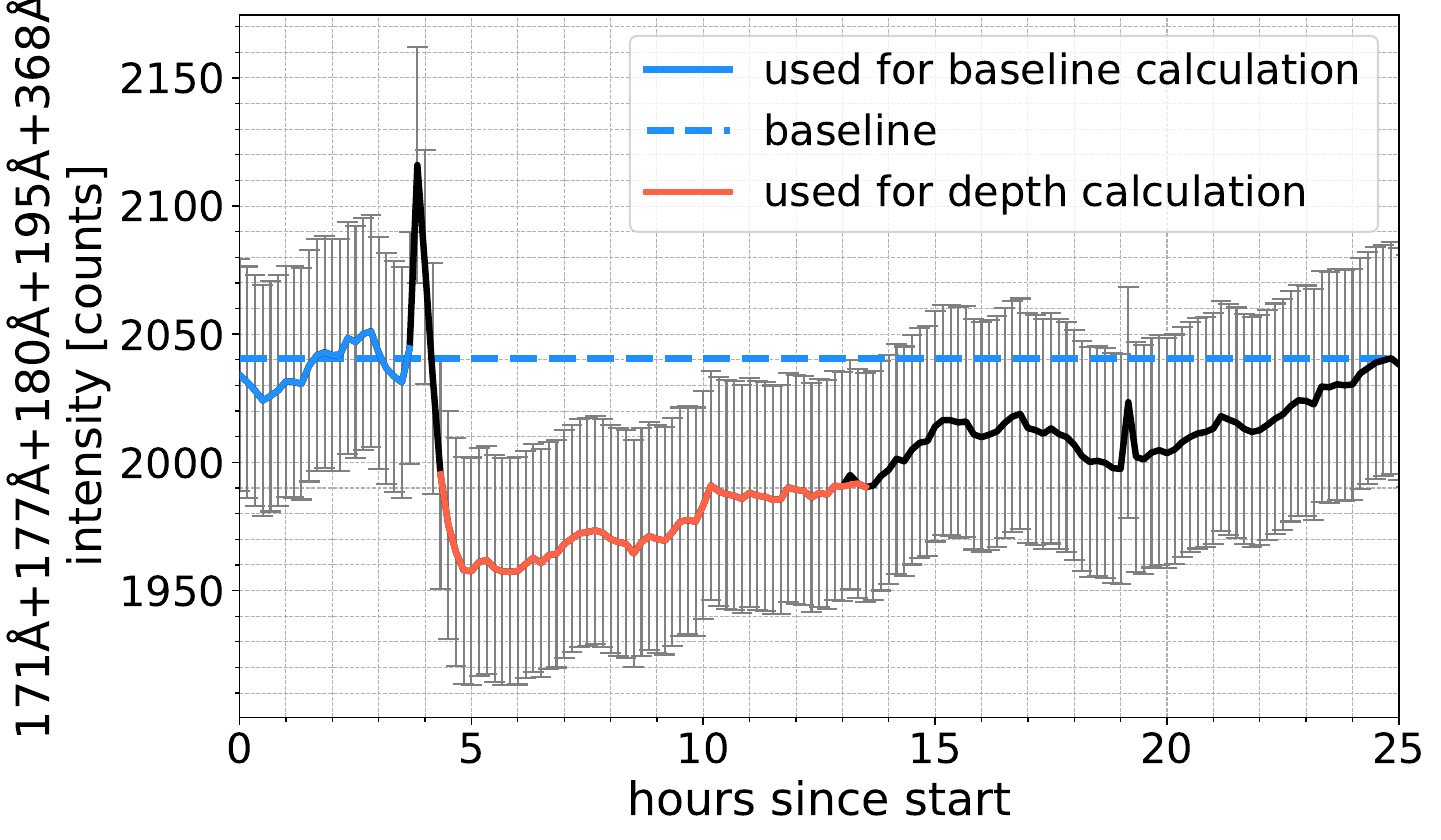}
\caption{Example light curve using ESCAPE effective area for 600-second integration time. Error bars include counting statistics, detector background noise, and scattered light. This light curve corresponds to the red curve in the bottom-middle plot in Figure \ref{fig:escape_light_curves}.
\label{fig:escape_single_light_curve}}
\end{figure}

\section{Results} \label{sec:results}

Using the methods described in prior sections, we can compare how the dimming detectability scales with stellar brightness and ISM-attenuation for both EUVE and ESCAPE. We considered values ranging from log$_{10}$\textit{F} of -10.5 to -13.5 erg~s$^{-1}$~cm$^{-2}$ and log$_{10}$ N(HI) of 17.5 to 19.0 cm$^{-2}$, which encompasses the actual values for well over 1000 stars \citep{Freund2022, Wood2005, Youngblood2025}. The results are shown in Figures \ref{fig:euve_detections} and \ref{fig:escape_detections}. Each point represents the best detection across a set of the emission-line combinations (left) or bands (right). For instance, if the depth/$\sigma_{depth}$ for the 100-300 \AA\ band was higher than that for the 150-250 \AA\ band, then that point gets represented in the figure. The emission lines considered for this analysis were chosen as known solar dimming-sensitive lines spanning roughly ambient-coronal temperatures ($\sim$0.5-2 MK): \ion{Fe}{9} 171~\AA, \ion{Fe}{10} 177~\AA, \ion{Fe}{11} 180~\AA, \ion{Fe}{12} 195~\AA, \ion{Fe}{13} 202~\AA, \ion{Fe}{14} 211~\AA, \ion{Mg}{9} 368~\AA, \ion{S}{14} 445~\AA, and \ion{Ne}{7} 465~\AA. The bands considered were chosen for spanning the wavelength regimes of the emission lines: 80-180 \AA, 150-250 \AA, and 100-300 \AA.

\begin{figure*}
    \plottwo{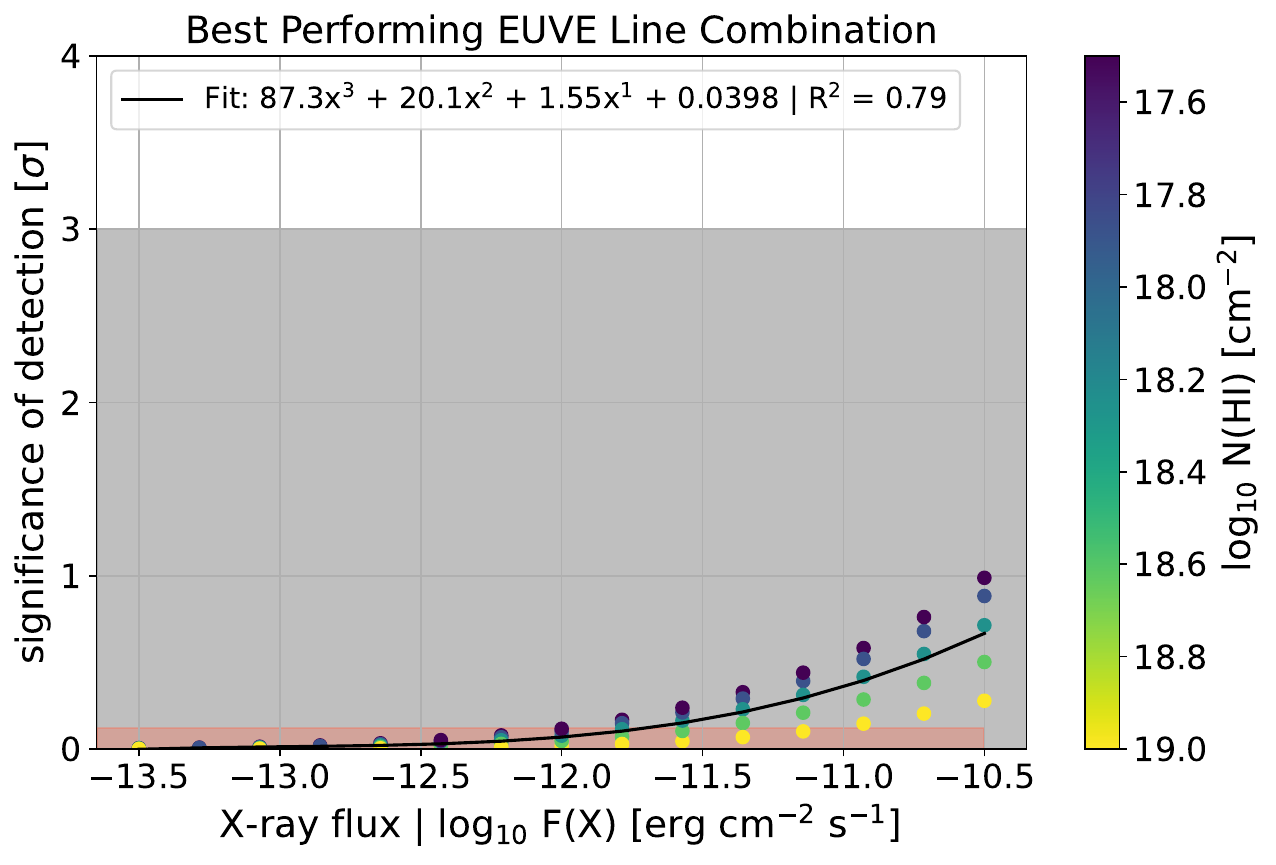}{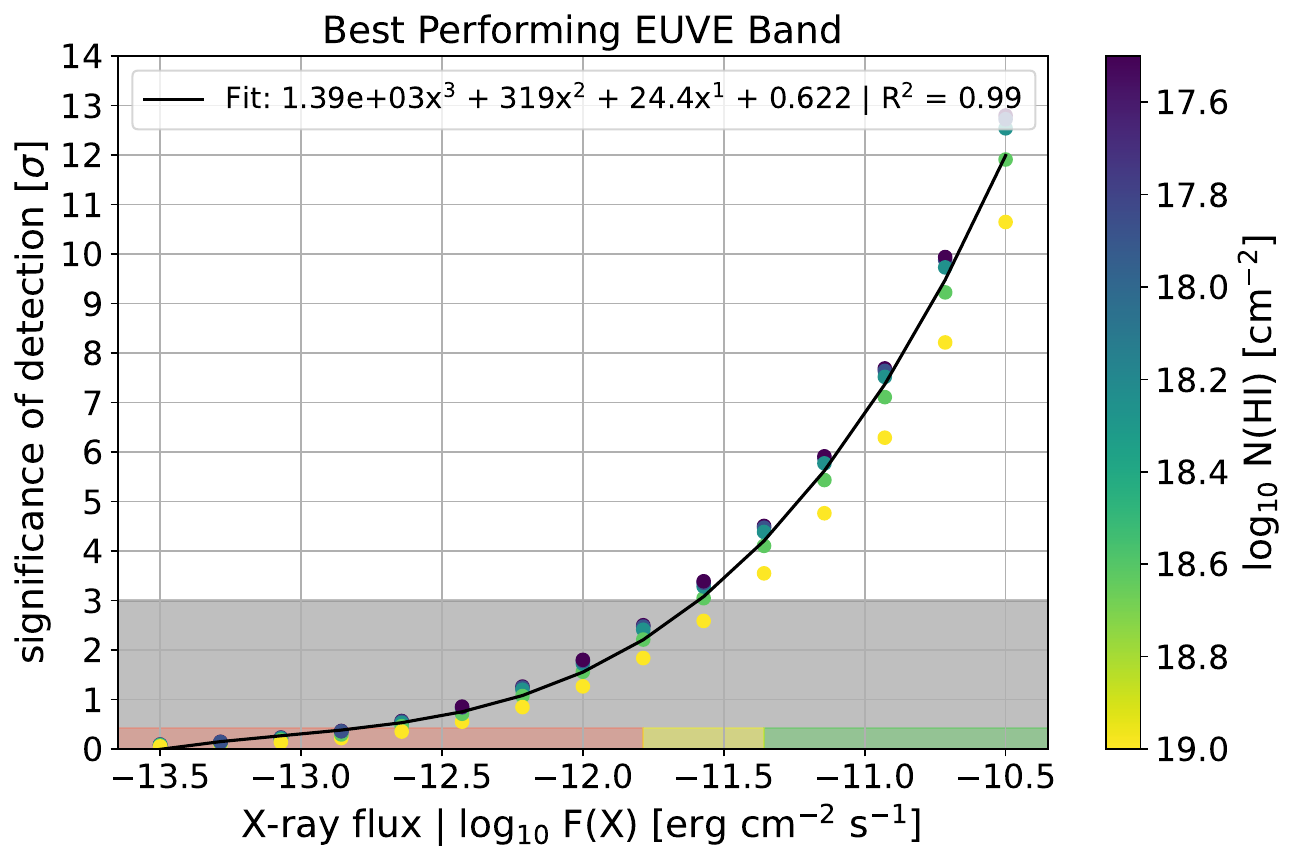}
    \caption{Dimming detectability for the EUVE instruments. Note that the line combinations can only be done with the EUVE DS/S spectrometers (left; low effective area) while the EUVE DS/S imager has higher effective area but is broad band (right). Each point in the plot corresponds to the best 5-line combo / band detection significance (vertical axis) for the associated X-ray flux (horizontal axis) and ISM column density (colorbar). The 5 discrete ISM column densities used are 5 evenly-spaced points between 17.5 and 19.0 cm$^{-2}$. The grey shaded area indicates no significant detection, that is $\leq 3 \sigma$. The red, yellow, green shaded bars at the bottom are described by the enumerated list in the body of the text.
    \label{fig:euve_detections}}
\end{figure*}

\begin{figure*}
    \plottwo{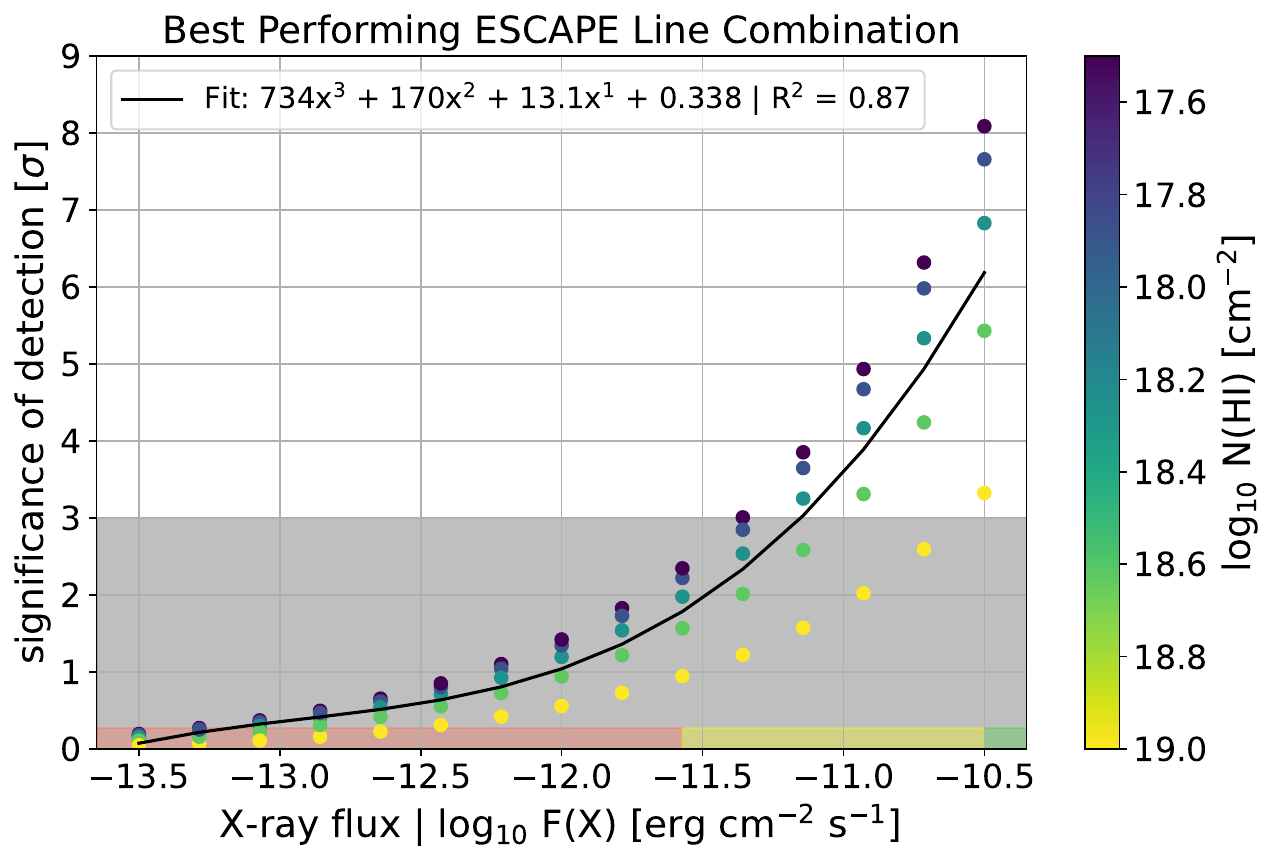}{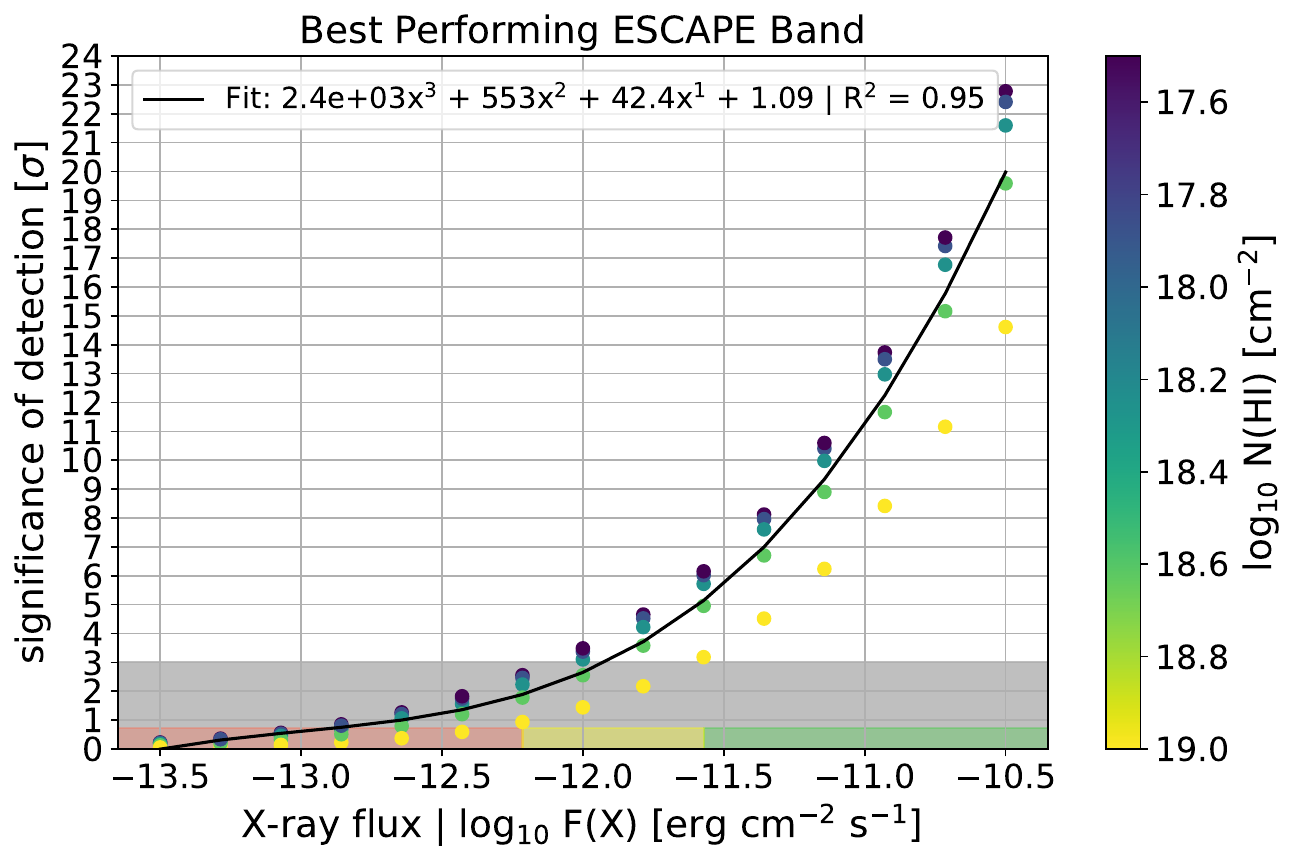}
    \caption{Same as Figure \ref{fig:euve_detections} but for ESCAPE. Lines or bands can be chosen post facto as these only represent different integrals across the spectrum measured by the single instrument, but here the same set of lines and bands were considered for both EUVE and ESCAPE.
    \label{fig:escape_detections}}
\end{figure*}

The relationship between dimming detectability and brightness is well fit by a third order polynomial (Figures ~\ref{fig:euve_detections} and \ref{fig:escape_detections}), which makes for a simple and convenient way to estimate how detectable dimming from any star might be \textit{given only its X-ray brightness}. The method does not distinguish the source of the EUV brightness, meaning that this technique is equally valid for stars with modest intrinsic EUV luminosity but are close to the Sun, or more distant stars with intrinsically large EUV luminosities (e.g., young or active stars). The spread around the fit comes from the ISM attenuation. By setting a minimum threshold on the acceptable significance of detection (3$\sigma$ here), three critical regimes can be defined: 

\begin{enumerate}
    \item For stars with X-ray fluxes above a particular value, $\sim$no amount of ISM attenuation is sufficient to reduce the dimming detectability below the threshold, e.g., in Figure \ref{fig:escape_detections} (right) this corresponds to an X-ray flux of -11.6, shaded green at the bottom of the plot if any points meet this criteria
    \item For stars with X-ray fluxes below a particular value, even the smallest amount of ISM attenuation is too much for the dimming to be detectable, e.g., X-ray flux of -12.2 in Figure \ref{fig:escape_detections} (right), shaded red
    \item The in-between case: dimming detectability is strongly dependent on the amount of ISM attenuation, e.g., for X-ray fluxes between -12.2 and -11.6 in Figure \ref{fig:escape_detections} (right), shaded yellow
\end{enumerate}

It is important to note that results presented here are all for a 600-second integration time and the line combinations are fixed at co-adding 5 lines. We did explore other values for these parameters but this combination of parameter space represents the most favorable we are aware of for EUV dimming detections. That exploration confirmed some important trade offs to keep in mind. Longer integration times improve the counting statistics and thus $\sigma_{depth}$, but they also flatten the dimming profile which reduces the estimated depth. Similarly, co-adding more lines improves counting statistics and thus $\sigma_{depth}$, but including lines with formation temperatures different than the quiescent coronal temperature (which therefore do not dim significantly) reduces the depth\footnote{The same is also true for broadband integrations.}. The upshot is that we now have a tool that can be easily run for any combination of emission lines, any spectral bandpass, and at any integration time\footnote{The code is publicly available at \citet{Mason2025-StellarDimmingCode}}; for photon-counting instruments like ESCAPE that would allow us to optimize these parameters to achieve the best balance of detectability with spectral and temporal resolution for each stellar target even \textit{after} the observations were made. The flux and column density are effectively fixed parameters for each stellar target, so results like those shown in Figure \ref{fig:escape_detections} can help make target selections. 

\section{Discussion} \label{sec:discussion}

There is strong evidence that stellar coronal dimming \textit{can} be detected, as suggested by the results in Figures \ref{fig:euve_detections} and \ref{fig:escape_detections}. These findings also illustrate why detections have been rare so far: achieving them requires a mission specifically optimized for this measurement. The EUVE DS/S spectrometers, for instance, lacked sufficient effective area, placing all of its data points in the non-detection region (shaded gray) in Figure \ref{fig:euve_detections} (left). Although the EUVE DS/S imager offered enough effective area, irradiance coronal dimming had not yet been discovered and thus the mission wasn't optimized specifically for its detection, resulting in only the one tantalizing but tentative case reported by \citet{Veronig2021}.

There are limitations to this study and the EUV dimming detection method generally. Our tool and the resultant analysis presented here assumes that the EUV and SXR fluxes are well correlated, an assumption that should be a relatively safe one given the reasoning and prior studies described in Section \ref{sec:adapting}. 
Additionally, the dimming profile analyzed here was one without many flares interfering with the measurement. \citet{Mason2019} performed a large statistical study of solar dimmings and found that this issue is easily overcome in determining the pre-event baseline by simply taking a median; however, in the dimming phase our algorithm takes a mean in order to more rapidly reduce the error bars than occurs when taking a median, and therefore the average is sensitive to flare peaks. Other algorithms could conceivably be developed to ignore flares in the dimming phase. Most flares tend to be short in duration (minutes) while dimming tends to last hours, but there are exceptions (e.g., EUV late phase, \citealt{Woods2011}) and if the number of flares is especially high, the irradiance may never be allowed to settle down to the CME-induced dimming level. However, \cite{Veronig2021} analyzed broad-band EVE Sun-as-a-star observations and found that the average time difference between flare peak and dimming start is $0.9 \pm 0.6$ hours, but for stellar cases the time separation was $>$2 hours, suggesting that at least in some cases the flares and dimmings may be \textit{more} temporally distinct on other stars. 

There are also other, complementary methods that could be employed to detect stellar CMEs. For example, if the eruption is moving toward or away from the observer, there should be a corresponding Doppler shift in the spectra and there are numerous reports of blueshifts of chromospheric lines during stellar flares (e.g., \citealt{Leitzinger2014, Namekata2022, Namekata2024}). Type II radio bursts would also be great indicators of stellar CMEs, but to date the searches have turned up empty (e.g., \citealt{Crosley2018a, Crosley2018b, Villadsen2019}). 

\section{Conclusion} \label{sec:conclusion}

If we want to know whether any particular exoplanet is a good candidate to harbor life, we need to know if the host star is hospitable. If the space climate is especially harsh, it may not be possible for planets to retain their atmospheres, which in turn greatly reduces the likelihood for there to be life as we know it there. This is not a hypothetical condition; recent studies with the James Webb Space Telescope have found that many terrestrial planets around M dwarfs have lost their atmospheres \citep{Lim2023, Mansfield2023, Xue2024, Scarsdale2024}. Looking ahead, atmospheric erosion on planets around more massive stars will have a direct impact on the most promising targets for the Habitable Worlds Observatory.

While we have been able to detect stellar flares for decades, stellar coronal mass ejections remain elusive. They are difficult to detect \citep{Leitzinger2022, Osten2021IAU}. The traditional method for the Sun is to use a coronagraph to directly image the CME, but that is many orders of magnitude harder to do for other stars.
There are a handful of other potential methods, including the EUV coronal dimming studied here that was a natural outgrowth of discovering that CME-induced dimming was detectable in solar EUV irradiance (that is, Sun-as-a-star) measurements. This discovery was only made about a decade ago. In that time, the community has combed through the existing archives of stellar data in EUV and adjacent wavelengths and enticing measurements have been made. But all of the missions/instruments to date have been optimized for other science goals. What's needed to unambiguously make these detections is instrumentation fit for purpose. 

These observations need to be made from space because Earth's atmosphere absorbs all of the relevant light. Single targets need to be observed for long durations (multiple days at minimum) to establish good pre-event baselines, capture the complete event profile, and increase the number of events observed. The effective area needs to be at least as high as the best we've flown to date, and ideally several times higher in order to get good signal to noise ratios and increase the number of good potential targets. Finally, while not a strict requirement, it is extremely beneficial if spectral and time integrations can be done as part of the data production because no a-priori guesses at the optimal values need to be made for each individual target. Such concepts already exist and there are surely other implementations that could achieve the same goals. Now all that is left is to see them fly.

\begin{acknowledgments}
\end{acknowledgments}

\appendix
\section{Impact of Mission Observing Efficiency} \label{sec:observing_efficiency}

Satellites in low Earth orbit (LEO) complete an orbit approximately every 90 minutes. For a mission attempting to observe a particular stellar target, interruptions may occur as the Earth or Moon obstruct the line of sight. These periodic disruptions define the spacecraft's observing efficiency, the fraction of time during which observations can proceed without interruption. For many space-based observatories in LEO, a reasonable observing efficiency is around 78\%, meaning that roughly 22\% of each 90-minute orbit is lost due to obstructions and the time it takes to reacquire the target.

\begin{figure}
\plotone{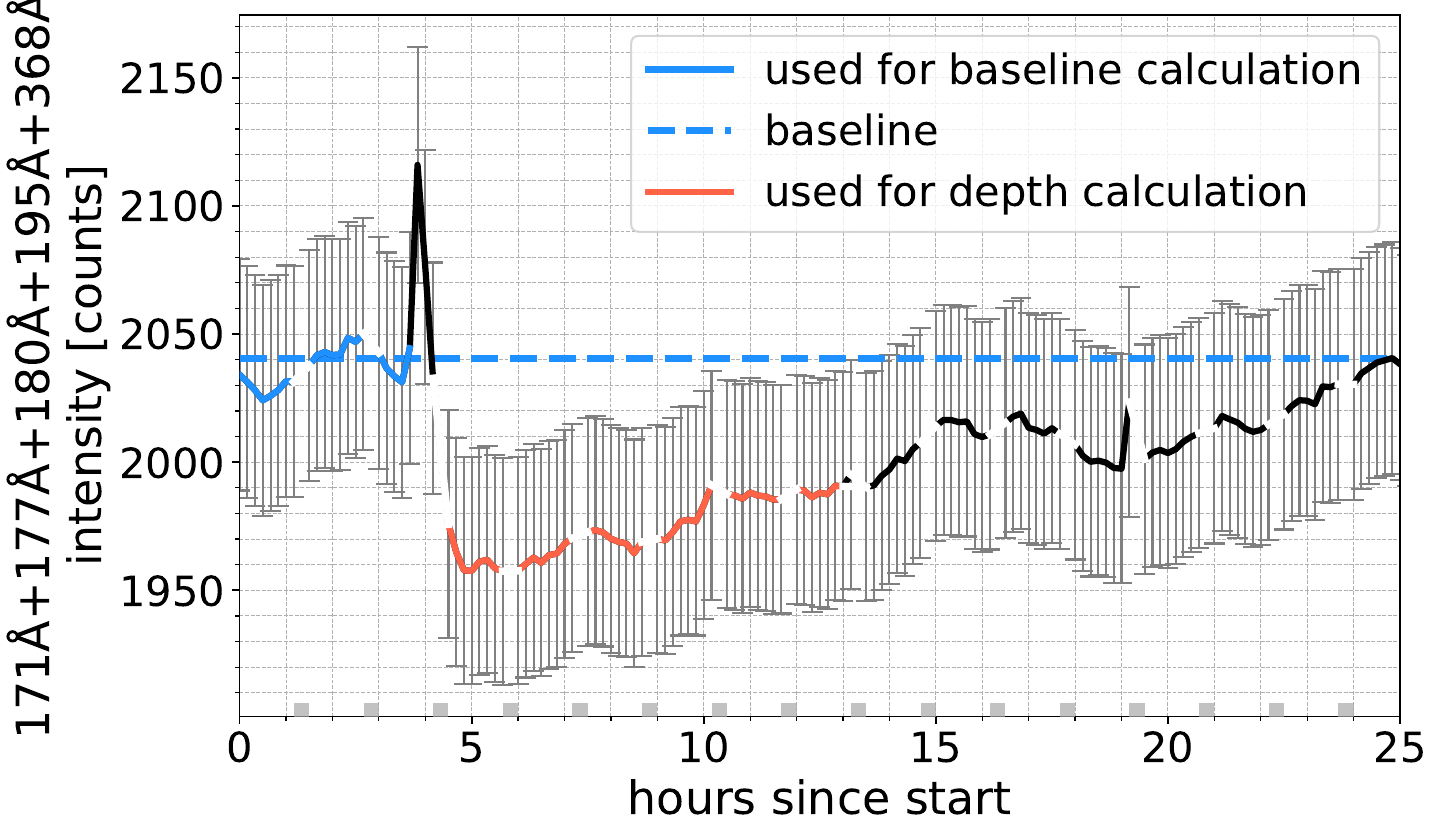}
\caption{Same as Figure \ref{fig:escape_single_light_curve} but with 78\% observing efficiency and a 90-minute orbit: a 20-minute period is dropped every 90-minutes. The difference from Figure \ref{fig:escape_single_light_curve} is most apparent when looking at the short gaps between the error bars; grey bars were added to the bottom of the plot to accentuate these periods. 
\label{fig:escape_single_light_curve_observing_efficiency}}
\end{figure}

This periodic interruption can be accounted for in the coronal dimming light curve analysis. To model this effect, we introduce artificial observing gaps into our simulated light curves, mimicking the expected cadence of a real mission in orbit (Figure \ref{fig:escape_single_light_curve_observing_efficiency}). However, because coronal dimmings typically evolve over timescales of several hours—such as the nine-hour duration of the event analyzed in this work—the overall impact of these interruptions remains minimal. The gradual nature of the dimming ensures that the essential feature of the light curve -- the trough highlighted in red in Figure \ref{fig:escape_single_light_curve_observing_efficiency} and described in Section \ref{sec:detection} -- remain detectable even when factoring in these periodic gaps in observation. Thus, there is no impact to the downstream analysis presented here, provided the observing efficiency does not dip to an egregiously low value. Additionally, even in LEO, it is possible to choose targets such that there will be no interruptions to observation. The details depend on the particulars of the spacecraft's orbit, time of year, and the list of available targets.

\vspace{5mm}
\facilities{SDO(EVE), SDO(AIA), EUVE}

\software{Analysis code for this paper \citep{Mason2025-StellarDimmingCode}, 
          Code for plots in this paper \citep{Mason2025-StellarDimmingPaperPlots},
          Data for this paper \citep{Mason2025-stellareuvdataset},
	   astropy \citep{Price-Whelan2018}, 
          ChatGPT \citep{chatgpt2020},
          IDL, 
          matplotlib \citep{Hunter2007}, 
          numpy \citep{Oliphant2006}, 
          scipy \citep{Scipy2020},
          SolarSoft \citep{SolarSoft2012}
          }

\bibliography{references}{}
\bibliographystyle{aasjournal}



\end{document}